\documentclass[12pt]{article}
\usepackage{graphicx}
\usepackage{amssymb}
\usepackage{amsmath}
\begin{document}

\title{$Z_3$-graded colour Dirac equations for quarks, 
 confinement  \\   and generalized Lorentz symmetries}

\author{Richard Kerner$^a$ and Jerzy Lukierski$^b$
\\
$^a$ Laboratoire de Physique Th\'eorique de la Mati\`ere Condens\'ee, 
\\
 Sorbonne-Universit\'e, 4 Place Jussieu, 75005 Paris, France;
\\
{e-mail: richard.kerner@upmc.fr}
\\  \\
 $^b$ Institute of Theoretical Physics, Wroc{\l}aw University, 
\\ Plac Maxa Borna 9,  Wroc{\l}aw, Poland
\\
{e-mail: jerzy.lukierski@ift.uni.wroc.pl}
}

\maketitle

%\vskip 0.2cm
{ Keywords: colour Dirac equation, quarks confinement, 
 $Z_3$ grading, \\ generalized Lorentz symmetries}

\begin{abstract}
We propose a modification of standard QCD description of the colour triplet of quarks
%describing quark fields endowed with colour degree of freedom 
by introducing a $12$-component colour generalization of Dirac spinor, with
built-in $Z_3$ grading playing an important algebraic role in quark confinement. In ``colour Dirac equations" 
the $SU(3)$ colour symmetry is entangled with
the $Z_3$-graded generalization of Lorentz symmetry, containing three $6$-parameter sectors related by $Z_3$ maps. 
The generalized Lorentz covariance requires simultaneous presence of $12$ colour Dirac multiplets,
which lead to the description of all internal symmetries of quarks: besides $SU(3) \times SU(2) \times U(1),$
the flavour symmetries and three quark families.
   \end{abstract}
\vskip 0.4cm
\noindent
{\bf{1. Introduction}} 
\vskip 0.3cm
\indent
It is well known that colour symmetries play a double role - they describe $SU(3)$ gauge symmetry group in QCD
and are linked with quark confinement, which is obtained usually as dynamical consequence of strong forces between quarks growing
linearly with their spatial separation (see e.g. \cite{Greensite}, \cite{WuHwang}). 

In the present paper we would like to propose an alternative algebraic approach to the
confining aspect of colour symmetries. For that purpose we replace the usual tensor product of Lorentz and colour 
$SU(3)$ group actions by the entanglement of space-time and colour symmetries generated by the $Z_3$ symmetry which plays also 
an important role in the appearance of fractional electric and baryonic charges of quarks.

We shall show that such an entanglement appears naturally when we generalize the derivation of the $4$-component Dirac equation
as given by particular $Z_2 \times Z_2$ symmetric coupling of a pair of $2$-component Pauli spinors (\cite{Pauli1926}), to the 
$Z_3 \times Z_2 \times Z_2$ symmetry which unifies in a specific manner (see (\ref{systemsix})) the system of six linear equations for 
six Pauli spinors. In such a way we obtain a new description of quarks endowed with colour as irreducible $12$-component analogs 
of Dirac spinors, with internal (colour) and space-time degrees of freedom entangled in a non-trivial manner.

By studying the solutions of free colour Dirac equation one arrives at possible algebraic explanation of quark 
confinement phenomenon: all exponential solutions of this system, which are wave-like, depend on complex wave vectors with imaginary parts damping
the free propagation of coloured quarks; however, certain 
cubic products of such solutions result in the cancellation of imaginary parts and produce propagating waves corresponding to the freely moving 
composite particle states (see e.g. \cite{Kerner2017}, \cite{Kerner2018b}, \cite{Kerner2019}).
\\
The plan of our paper is as follows: 

In Section 2 we present how to obtain the $12$-component colour
Dirac equation which describes the dynamics of quark and anti-quark endowed with colour and spin by introducing 
$12$-component colour Dirac field equations displaying the $Z_3 \times Z_2 \times Z_2$ symmetry.

In Section 3 we consider the extension of relativistic symmetry by incorporating in $18$-parameter
generalized Lorentz symmetry the standard $6$-parameter Lorentz subgroup and the additional pair of complex-conjugated $6$-parameter
Lorentz-like sectors.
%so that the resulting 
Such $Z_3$-generalization ${\cal{L}}$ of the Lorentz algebra can be decomposed into the following $Z_3$-graded sum of three sectors:
\begin{equation}
{\cal{L}} = L^{(0)} \oplus L^{(1)} \oplus L^{(2)}, \; \; \; 
\left[ L^{(k)}, L^{(m)} \right] \in L^{(k+m)},  
\label{ThreeL}
\end{equation}
where $k, m = 0,1,2$ and $(k+m)$ is mod $3$, $L^{(0)}$ 
describes the standard Lorentz sector, while adding $L^{(1)}$ and 
$L^{(2)} =\left( L^{(1)} \right)^{\dagger}$ (${\dagger}$ denotes Hermitian conjugate)  
extends it to a $Z_3$-graded generalized Lorentz algebra. 

To obtain the representation of generalized Lorentz  algebra 
$\cal{L}$ one should introduce the set of twelve $12 \times 12$ generalized Dirac matrices 
$\Gamma^{\mu}_{F} = \left( \Gamma_{F}^0, \; \Gamma_{F}^k \right)$ 
$ (F=1,2,..12), $ 
%describing $12$  entangled colour quark and anti-quark fields. 
\noindent
In order to show it we introduce 12$\times$12  matrix $\cal{S}$
 describing the spinor representation of the generalized Lorentz algebra ${\cal{L}}$, 
 which contains a $6$-parameter subgroup
${\cal{S}}^{(0)} \subset {\cal{S}}$ representing the standard Lorentz group.

We shall study  the transformations of colour Dirac matrices $\Gamma^\mu_{(F)}$
under the $18$-parameter spinor transformation $\Psi' = {\cal{S}} \Psi$, where $\Psi$ is a 12-component 
  colour Dirac spinor, and includes its standard Lorentz subgroup $\Psi' = {\cal{S}}^{(0)} \Psi $. 
 We obtain that the covariance under spinor Lorentz transformations ${\cal{S}}^{(0)} \Gamma^{\mu}_{(F} [{\cal{S}}^{(0)}]^{-1}$ 
requires the introduction of Lorentz doublets of colour-Dirac  matrices, 
while the closure of the map ${\cal{S}} \Gamma^{\mu}_{(F)} {\cal{S}}^{-1}$ leads to the appearance 
of  $12$ different Lorentz doublets of 
matrices $\Gamma^{\mu}_{(F)}$. One can further argue that the lowest-dimensional spinor space 
on which  act the generalized Lorentz transformations  in a closed and faithful way describes six different types 
of coloured quarks. The standard Lorentz covariance and the presence of Lorentz doublets of colour Dirac fields 
leads to an additional $Z_2$-factor 
in front of $Z_3 \otimes Z_2 \otimes Z_2$, which may be responsible for the appearance of weak isospin doublets of quarks.
In this way extending $Z_3$ to $Z_2 \times Z_3$ permits
to introduce the Standard Model's internal symmetries $SU(3) \times SU(2) \times U(1)$, which by gauging generate the  
octet of gluons, three vector mesons $B^a_{\mu}$ and the electromagnetic field $A_{\mu}$. Finally, the generalized Lorentz
covariance implies the six-fold enlargement of the representation space of standard Lorentz symmetries and allows
 to accomodate quark flavour doublets and the triplet of quark families. 

In Section $4$ we display the complete basis of solutions of the generalized "colour" Dirac equation, all of which are represented 
by exponential with non-vanishing damping factors. Further, we illustrate the confinement by showing that 
certain ternary products of such damped solutions can propagate freely, and can asymptotically represent the composite free baryon state.

 The  colour Dirac spinor components satisfy 
sixth order homogeneous field equation, which in the four-momentum space factorizes into three mass
shells: one with real mass and two with conjugate complex masses, related respectively with the $Z_3$-graded sectors 
$L^{(0)}, \;  L^{(1)}$ and $L^{(2)}$ of ${\cal{L}}$. We point out that such a triplet of masses
  can coincide with the mass spectrum of a particular $Z_3$-covariant perturbative Lee-Wick QFT \cite{LeeWick}. 

We conclude that recent results in the description of renormalizability and unitarity of the Lee-Wick perturbative QFT 
\cite{AnselmiPiva} justify the conjecture that our model of coloured quarks may be also renormalizable and unitary.   
\vskip 0.3cm
\noindent
{\bf {2. From Dirac to coloured Dirac equation.}} 
\vskip 0.2cm
\indent
The Dirac equation for the electron (or any spin $\frac{1}{2}$ particle with non-zero mass $m$) 
(\cite{Dirac1928})
 can be written in a compact way as follows:
%, described by four-component
%Dirac bispinor $\psi = [ \psi_{+} , \psi_{-} ]^T$) 
%can be written in a compact way as follows:
\begin{equation}
\gamma^{\mu} \; p_{\mu} \; \psi = mc \; \psi \; \; \; {\rm with} \; \; \psi = (\psi_{+}, \psi_{-} )^T  \; , 
\label{Dirac1}
\end{equation} 
where $p_{\mu}\!\! =\!\! - i \hbar \partial_{\mu}$, $\psi_{\pm}$ are two complex $2$-component Pauli spinors,
and as Dirac matrices $\gamma^{\mu}$ one can choose
\begin{equation}
\gamma^0\!\! =\!\! \sigma_3 \otimes {\mbox{l\hspace{-0.55em}1}}_{2}, \;  
\gamma^k\!\! =\!\! (i \sigma_2) \otimes \sigma^k,
\label{gammasdef}
\end{equation}
where $\sigma_0 = {\mbox{l\hspace{-0.55em}1}}_{2}$, and
$\sigma^k \; (k\!\! =\!\! 1,2,3)$ are Pauli matrices.
The Dirac matrices realize the $4$-dimensional Clifford algebra 
\begin{equation}
\gamma^{\mu} \gamma^{\nu} + \gamma^{\nu} \gamma^{\mu}\!\! =\!\! 2 \; \eta^{\mu \nu} \; {\mbox{l\hspace{-0.55em}1}}_{4}, \; \; \; 
\eta^{\mu \nu}\!\! =\!\!  {\rm diag} (+,-,-,-).
\label{Clifford}
\end{equation}
Under the Lorentz transformation
\begin{equation}
x^{\mu}\! \rightarrow\! x^{\mu'}\!\! =\!\! \Lambda^{\mu'}_{\; \; \nu} \; x^{\nu}
\label{Lorentz0}
\end{equation}
the spinor field $\psi\!\! =\!\! \psi^{A} \; (A\!\! =\!\! 1,2,3,4)$ in (\ref{Dirac1}) transforms as follows:
\begin{equation}
\psi' (x^{\rho'}) = \psi' (\Lambda^{\rho'}_{\; \; \mu} x^{\mu}) = S \psi (x^{\mu}) \; .
\label{spintransf}
\end{equation}
In order to ensure the standard Lorentz covariance, the condition 
relating the vectorial and spinorial realizations of the Lorentz group $O(3,1) \simeq SL(2, {\bf C})$ is:
\begin{equation}
S \gamma^{\mu'} S^{-1} = \Lambda^{\mu'}_{\; \; \nu} \gamma^{\nu}\; .
\label{Lorentz3}
\end{equation}
The spinorial representation $S$ is given by the formula
\begin{equation}
S = exp \left( - \frac{i}{4} \omega_{\mu \nu} \sigma^{\mu \nu} \right),
\label{spinS}
\end{equation}
where $\sigma^{\mu \nu}=\frac{i}{2} [ \gamma^{\mu} , \gamma^{\nu} ]$, and the corresponding
infinitesimal vectorial representation is given by the formula 
\begin{equation}
\Lambda^{\mu}_{\; \; \nu} = \delta^{\mu}_{\; \; \nu} + \omega^{\mu}_{\; \; \nu}, \; {\rm where} \; 
\omega_{\mu \nu} = \eta_{\mu \lambda} \;  \omega^{\lambda}_{\; \; \nu}  = - \omega_{\nu \mu}. 
\label{parameters}
\end{equation}
with three independent Lorentz boosts
($\omega_{0k} = - \omega_{k0}$) and three independent spatial rotations ($\omega_{ij} = - \omega_{ji}$)).

The generalized Dirac equation incorporating colour degrees of freedom in a $Z_3$-symmetric way was proposed in
\cite{Kerner2017}, \cite{Kerner2018b}, \cite{Kerner2019} after introducing three pairs of independent  Pauli  spinors
{\small 
$$\varphi_{+} = \begin{pmatrix} \varphi_{+}^1 \cr \varphi_{+}^2 \end{pmatrix}, \; \; 
\varphi_{-} = \begin{pmatrix} \varphi_{-}^1 \cr \varphi_{-}^2 \end{pmatrix}, \; \;  
\chi_{+} = \begin{pmatrix} \chi_{+}^1 \cr \chi_{+}^2 \end{pmatrix},  $$
\begin{equation}
 \chi_{-} = \begin{pmatrix} \chi_{-}^1 \cr \chi_{-}^2 \end{pmatrix}, \; \; 
\psi_{+} = \begin{pmatrix} \psi_{+}^1 \cr \psi_{+}^2 \end{pmatrix}, \; \; 
\psi_{-} = \begin{pmatrix} \psi_{-}^1 \cr \psi_{-}^2 \end{pmatrix}.
\label{colored}
\end{equation}}
with Pauli sigma-matrices acting on them in a natural way. 
These three Pauli spinors $\varphi_{+}, \; \chi_{+}$ and $\psi_{+}$
are conventionally named ``red", ``blue" and ``green", while their antiparticle counterparts 
$\varphi_{-}, \; \chi_{-}$ and $\psi_{-}$ are called, respectively, ``cyan", ``yellow" and ``magenta".

The cyclic group $Z_3$ is represented on the complex plane by multiplicative group of three complex numbers,
generated by powers of $j = e^{\frac{2 \pi i}{3}}$, namely:
\begin{equation}
 j= e^{\frac{2 \pi i}{3}}, \; \; j^2 = e^{\frac{4 \pi i}{3}}, \; \;\; j^3 = 1, \; \; \; 1 + j + j^2 = 0.
\label{Z3group}
\end{equation}

The $Z_2 \times Z_2$ symmetry of the Dirac equation can be made explicit if we 
multiply (\ref{Dirac1}) by $\gamma^0$ and get a system of two equations entangling two Pauli  spinors:
\begin{equation}
\begin{split}
& E \; \psi_{+} = mc^2 \; \psi_{+} + c {\boldsymbol{\sigma}} {\bf p} \; \psi_{-},
\\
%\begin{equation}
&E \; \psi_{-} = - mc^2 \; \psi_{-} + c {\boldsymbol{\sigma}} {\bf p} \; \psi_{+}\; .
\end{split}
\label{DiracPauli}
\end{equation} 

The system (\ref{DiracPauli}) displays two discrete $Z_2$ symmetries: the space reflection
simultaneously changes directions of spin and momentum, 
${\boldsymbol{\sigma}} \rightarrow - {\boldsymbol{\sigma}}, \; \; \; {\bf p} \rightarrow - {\bf p},$
and the particle-antiparticle symmetry realized  in (\ref{DiracPauli}) by the transformation  
$m \rightarrow - m, \; \psi_{+} \rightarrow \psi_{-}, \; \; \; \psi_{-} \rightarrow \psi_{+}.$

In what follows, we extend the $Z_2 \times Z_2$ symmetry by $Z_3$ group, so that the system
will mix not only the two spin $\frac{1}{2}$ states and particles with anti-particles, but the three colours as well.
The standard Dirac equation (\ref{DiracPauli}) 
%expressed in terms of two entangled Pauli spinors $\psi_{\pm}$ in(\ref{DiracPauli})  
is extended in the following way in terms of six entangled Pauli spinors:
% (\ref{colored}):
\begin{equation}
\begin{split}
& E \; \varphi_{+} = mc^2 \, \varphi_{+} + c \; {\boldsymbol{\sigma}} \cdot {\bf p} \, \chi_{-},
\\
&E \; \varphi_{-} = - mc^2 \, \varphi_{-} + c \; {\boldsymbol{\sigma}} \cdot {\bf p} \, \chi_{+}
\\
&E \; \chi_{+} = j \; mc^2 \, \chi_{+} + c \; {\boldsymbol{\sigma}} \cdot {\bf p} \, \psi_{-},
\\
&E \; \chi_{-} = - j \; mc^2 \, \chi_{-} + c \; {\boldsymbol{\sigma}} \cdot {\bf p} \, \psi_{+}
\\
&E \; \psi_{+} = j^2 \;  mc^2 \, \psi_{+} + c \; {\boldsymbol{\sigma}} \cdot {\bf p} \, \varphi_{-},
%\begin{equation}
\\
&E \; \psi_{-} = -j^2 \;mc^2 \, \psi_{-} + c \; {\boldsymbol{\sigma}} \cdot {\bf p} \, . \varphi_{+}
\end{split}
\label{systemsix}
\end{equation}
The particle-antiparticle $Z_2$-symmetry appears as $m \rightarrow -m$ and simultaneously 
$(\varphi_{+}, chi_{+}, \psi_{+} ) \rightarrow (\varphi_{-}, \chi_{-}, \psi_{-})$ and vice versa;
the $Z_3$-colour symmetry is realized by multiplication of mass $m$ by $j$ each time the colour changes,
i.e. more explicitly, $Z_3$ symmetry is realized as follows:
\begin{equation}
m \rightarrow jm, \; \; \; \varphi_{\pm} \rightarrow \chi_{\pm} \rightarrow \psi_{\pm} \rightarrow \varphi_{\pm},
\label{Z3first}
\end{equation}
\begin{equation}
m \rightarrow jm, \; \; \; \varphi_{\pm} \rightarrow \psi_{\pm} \rightarrow \chi_{\pm} \rightarrow \varphi_{\pm},
\label{Z3second}
\end{equation}
 
The energy operator is diagonal;
%and its action on the spinor-valued $12$-component column 
%can be represented as a $6 \times 6$ ${\mbox{l\hspace{-0.55em}1}}_{2}$-valued unit matrix.
the mass operator is diagonal as well, but its elements are described by the powers of the 
sixth root of unity $q= e^{\frac{2 \pi i}{6}}: \; q^6 =1$, 
$q = - j^2, \; q^2 = j, \; q^3 = -1, \; q^4 = j^2, \; q^5 = - j.$
Choosing a particular basis in the space of
``coloured spinors" (\ref{colored}), such that $\Psi^T = [ \varphi_{+}, \varphi_{-}, \chi_{+}, \chi_{-}, \psi_{+}, \psi_{-} ],$
we rewrite (\ref{systemsix}) in compact form as a $12$-component equation:
\begin{equation}
{\small\hbox{$
E \; {\mbox{l\hspace{-0.55em}1}}_{3} \otimes {\mbox{l\hspace{-0.55em}1}}_{2} \otimes {\mbox{l\hspace{-0.55em}1}}_{2} \; \Psi \!\! =\!\!
\left[ Q_3 \otimes \sigma_1 \otimes c \; {\boldsymbol{\sigma}}\! \cdot\! {\bf p}\! +\!  mc^2  B \otimes \sigma_3
\otimes {\mbox{l\hspace{-0.55em}1}}_{2} \right]\! \Psi.$}}
\label{matDirac1}
\end{equation}
%where 
\begin{equation}
\; \; \; {\rm where} \; \; \; B = \begin{pmatrix} 1 & 0 & 0 \cr 0 & j & 0 \cr 0 & 0 & j^2 \end{pmatrix}
\; \; \; \; Q_3 = \begin{pmatrix} 0 & 1 & 0 \cr 0 & 0 & 1 \cr 1 & 0 & 0 \end{pmatrix}\; .
\label{BandQ3}
\end{equation} 
Note that $B^2 = B^{\dagger}$, with $B^{\dagger}$ the Hermitian conjugate of $B$, $B^3 = {\mbox{l\hspace{-0.55em}1}}_{3}$,
$Q_3^2 = Q_3^{\dagger}$ and $Q_3^3 = {\mbox{l\hspace{-0.55em}1}}_{3}$. 
Eq. (\ref{matDirac1}) can be presented in a way recalling much better the structure of the original Dirac equation (\ref{Dirac1})
if we multiply the equation (\ref{matDirac1}) from the left by $B^{\dagger} \otimes \sigma_3 \otimes {\mbox{l\hspace{-0.55em}1}}_{2}$.
We get
\begin{equation}
\left[ E \; B^{\dagger} \otimes \sigma_3 \otimes {\mbox{l\hspace{-0.55em}1}}_{2} \!
-\! Q_2 \otimes (i \sigma_2) \otimes c {\boldsymbol{\sigma}}\! \cdot\! {\bf p}  \right] \Psi
\!\! =\!\! mc^2 \; {\mbox{l\hspace{-0.55em}1}}_{12}  \Psi.  
%\otimes {\mbox{l\hspace{-0.55em}1}}_{2} \otimes {\mbox{l\hspace{-0.55em}1}}_{2} \; \Psi.
\label{matDirac2}
\end{equation}
Here $B^{\dagger}  = {\rm diag} [1, j^2, j] $, so that $B B^{\dagger} = B^{\dagger} B = {\mbox{l\hspace{-0.55em}1}}_{3}$ 
and $Q_2 = B Q_3$. 
The full set of matrices $Q_A$ and $Q^{\dagger}_B, \; A, B = 1,2,3,$ together with two diagonal
traceless matrices $B$ and $B^{\dagger}$ generated by $B$ and $Q_3$ form a special basis of the $SU(3)$ algebra \cite{Kac1994}.
They can be obtained by iteration, using the following multiplication table: 
\begin{equation}
\begin{split}
& \; \; \; \; \; \; B Q_A\!\! =\!\! j^2 Q_A B\!\! =\!\! Q_{A+1}, \; \; B^{\dagger} Q_A\!\! =\!\! j Q_A B^{\dagger}\!\! =\!\! Q_{A-1},
\\
& \; \; \; \; \; \; Q^{\dagger} B\!\! =\!\! j^2 B Q^{\dagger}_A\!\! =\!\! Q^{\dagger}_{A-1}, \; \; 
Q^{\dagger}_A B^{\dagger}\!\!=\!\! B^{\dagger} Q^{\dagger}_A\!\! =\!\! Q^{\dagger}_{A+1},
\\
%\begin{equation}
& \; \; \; \; \; \; \; \; \; Q_A Q_{A-1}\!\! =\!\! j Q^{\dagger}_{A+1}, \; \; Q^{\dagger}_{A-1} Q^{\dagger}_A\!\! =\!\! j^2 Q_{A+1},
\\
&Q_A Q_{A+1}^{\dagger} \! \! = \! \! B^{\dagger}, \; \; Q_A Q_{A-1}^{\dagger} \! \! = \! \! B, 
\; \; Q_A^{\dagger} Q_{A-1} \! \! = \! \! j B^{\dagger}, \; \;
Q_A^{\dagger} Q_{A+1} \! \! = \! \! j^2 B.
\end{split}
\label{RelBQ}
\end{equation}
and of course $Q_A Q_A^{\dagger} = Q_a^{\dagger} Q_A = {\mbox{l\hspace{-0.55em}1}}_{3}.$
%$$Q_A B^{\dagger} = j^2 Q_{A-1}, \; \; B^{\dagger} Q_A = Q_{A-1}, \; \; B Q_A = Q_{A+1}, \; \; Q_A B = j Q_{A+1},$$
%\begin{equation}
%B^{\dagger} Q_A = Q_{A-1}, \; \;  B^{\dagger} Q_A^{\dagger} = j^2 Q^{\dagger}_{A+1}, \; \; Q_A Q_{A-1} = j Q^{\dagger}_{A+1},
%\label{RelBQ}
%\end{equation}
where the indices $A, \; A+1, \; A-1$ are always taken {modulo} $3$, 
so that e.g. $3+1 \mid_{modulo \; 3} = 4 \mid_{modulo \; 3} = 1$, etc.,
and the cube of each of the eight matrices in (\ref{RelBQ}) is the unit $3 \times 3$ matrix.

Let us introduce the four $12 \times 12$ matrices $\Gamma^{0}, \; \Gamma^{i}$ given by the following formula (\ref{Gammasbig}).
\begin{equation}
\Gamma^0 = B^{\dagger} \otimes \sigma_3 \otimes {\mbox{l\hspace{-0.55em}1}}_{2}, \; \; \; \; \; 
\Gamma^k = Q_2 \otimes (i \sigma_2) \otimes \sigma^k   \; ,                                      
\label{Gammasbig}
\end{equation}
Now the system (\ref{matDirac2}) can be written in the $12$-dimensional Dirac-like form:
%concise form which suggests the possibility of implementing 
%the Lorentz covariance:
\begin{equation}
\Gamma^{\mu} \; p_{\mu} \; \Psi = mc \; \Psi, \; \; {\rm with} \; \; p_0 = E/c \; .
\label{Dirac3}
\end{equation}
It can be calculated \cite{Kerner2018b} that the ``colour Dirac operator" on the left-hand side of Eq. (\ref{Dirac3}) has the following important
algebraic properties:
%\begin{equation}
$$\left( \Gamma^{\mu} \; p_{\mu} \right)^6 = (p_0^6 - \mid {\bf p} \mid^6) \; {\mbox{l\hspace{-0.55em}1}}_{12}, \; \;  
{\rm det} (\Gamma^{\mu} \; p_{\mu}) = (p_0^6 - \mid {\bf p} \mid^6)^2,$$
\begin{equation}
p_0^6 -\! \mid\! {\bf p}\! \mid^6 = (p_0^2 -\! \mid\! {\bf p}\mid^2)(p_0^2 - j\! \mid\! {\bf p}\! \mid^2)(p_0^2 - j^2 \mid {\bf p}\mid^2).
\label{detGamma}
\end{equation}
%It is worthwhile to note that
%\begin{equation}
%p_0^6 - \mid {\bf p}\mid^6 = (p_0^2 - \mid {\bf p}\mid^2)(p_0^4 + p_0^2 \mid {\bf p}\mid^2 + \mid {\bf p}\mid^4)
%\label{Pproduct}
%\end{equation}
\vskip 0.3cm
\noindent
 {\bf {3. Implementing standard and generalized Lorentz  Covariance}} 
\vskip 0.2cm
\indent
It should be stressed that the $12 \times 12$ matrices $\Gamma^{\mu}$ appearing in the coloured Dirac equation
(\ref{Dirac3}) do not span 4-dimensional Clifford algebra. 
%as the $4 \times 4$ Dirac matrices $\gamma^{\mu}$. 
In fact, the $Z_3 \otimes Z_2$ structure 
of $\Gamma^{\mu}$-matrices implies that only their sixth powers are proportional to the unit matrix ${\mbox{l\hspace{-0.55em}1}}_{12}$
(see also (\ref{detGamma})).
Thus, in order to obtain the realization of $D=4$ Lorentz algebra generators 
one can not use just two standard commutators 
\begin{equation}
 J_i = \frac{i}{2} \; \epsilon_{ijk} \left[ \Gamma^j, \Gamma^k \right], \; \; \; K_l = \frac{1}{2} \; \left[ \Gamma_l, \Gamma_0 \right] \; .
\label{JotKa}
\end{equation} 
However, the generators $\left( J^{(0)}_i, \; K^{(0)}_l \right)$ satisfying the standard Lorentz algebra relations
(see also (\ref{modulocomm}) for $r=0, s=0$) can be defined by {triple} commutators:
\begin{equation}
\begin{split}
&\left[ J_i, \left[J_j, J_k \right] \right] = \left( \delta_{ij}\delta_{kl} - \delta_{ik} \delta_{jl} \right) \; J^{(0)}_l,
\\
%\begin{equation}
&\left[ K_i, \left[K_j, K_k \right] \right] = \left( \delta_{ij}\delta_{kl} - \delta_{ik} \delta_{jl} \right) \; K^{(0)}_l \; .
\end{split}
\label{triplecomm}
\end{equation}
Indeed, substituting in (\ref{triplecomm}) the explicit form of $\Gamma^{\mu}$ given in (\ref{Gammasbig}),
we get
\begin{equation}
\begin{split}
&J_i = - \frac{i}{2} \; Q_2^{\dagger} \otimes {\mbox{l\hspace{-0.55em}1}}_{2} \otimes \sigma_i, \; \; \; K_l = 
- \frac{1}{2} \; Q_1 \otimes \sigma_1 \otimes \sigma_l,
\\
%\begin{equation}
&J^{(0)}_i = - \frac{i}{2} \; {\mbox{l\hspace{-0.55em}1}}_{3} \otimes {\mbox{l\hspace{-0.55em}1}}_{2} \otimes \sigma_i, \; \; \; 
K^{(0)}_l = - \frac{1}{2} \; {\mbox{l\hspace{-0.55em}1}}_{3}  \otimes \sigma_1 \otimes \sigma_l.
\label{NewJK}
\end{split}  
\end{equation}
\indent
In order to close the generalized Lorentz algebra (\ref{ThreeL}) where 
$L^{(0)}\!\! =\!\! ( J^{(0)}_i, K^{(0)}_j ),$ \;   
$L^{(1)}\!\! =\!\! ( J^{(1)}_i, K^{(1)}_j ),$ \;   
$ L^{(2)}\!\! =\!\! ( J^{(2)}_i, K^{(2)}_j ),$ one should supplement (\ref{triplecomm})
by two missing triple commutators:
\begin{equation}
\begin{split}
&\left[ J_i, \left[J_j, K_k \right] \right] = 
\left( \delta_{ij}\delta_{kl} - \delta_{ik} \delta_{jl} \right) \; K^{(2)}_l,
\\
&\left[ K_i, \left[K_j, J_k \right] \right] = \left( \delta_{ij}\delta_{kl} - \delta_{ik} \delta_{jl} \right) \; J^{(1)}_l,
\end{split}
\label{tripcommextra}
\end{equation}
where using the representation (\ref{NewJK}) we get
\begin{equation}
J^{(1)}_l = -\frac{i}{2} \; Q_3 \otimes {\mbox{l\hspace{-0.55em}1}}_{2} \otimes \sigma_l, \; \; \; \; K^{(2)}_i 
= - \frac{1}{2} \; Q_3^{\dagger} \otimes \sigma_1 \otimes \sigma_i.
\label{J1K2}
\end{equation}
The full set of $Z_3$-graded relations defining the algebra (\ref{ThreeL}) ($r, s, \; r+s$ are modulo $3$):
\begin{equation}
\begin{split}
&\left[ J^{(r)}_i, J^{(s)}_k \right] \!\! =\!\! \epsilon_{ikl} J^{(r+s)}_l, \; \; 
  \left[ J^{(r)}_i, K^{(s)}_k \right]\!\! =\!\! \epsilon_{ikl} K^{(r+s)}_l, 
	\\
%\begin{equation}
 &\left[ K^{(r)}_i, K^{(s)}_k \right]\!\! =\!\! - \epsilon_{ikl} J^{(r+s)}_l.
\end{split}
\label{modulocomm}
\end{equation}
We see that from commutators $[ K_i^{(1)}, K_m ^{(1)} ] \simeq J^{(2)}$ and $[ J^{(1)}, J^{(1)} ] \simeq J^{(2)}$ one gets
the remaining generators of ${\cal{L}}$:
%the following extension of the formulae (\ref{J1K2}):
\begin{equation}
J^{(2)}_i = - \frac{i}{2} \; Q_3^{\dagger} \otimes {\mbox{l\hspace{-0.55em}1}}_{2} \otimes \sigma_i, \; \; \; 
K_m^{(1)} = - \frac{1}{2} \; Q_3 \otimes \sigma_1 \otimes \sigma_m.
\label{J2K1}
\end{equation}
The formulae (\ref{NewJK}, \ref{J1K2}) and ({\ref{J2K1}) describe the realization of ${\cal{L}}$ which follows from
the choice (\ref{Gammasbig}) of matrices $\Gamma^{\mu}$. 

Before considering standard and generalized Lorentz covariance we shall introduce the following notation:
\begin{equation}
\Gamma^{\mu}_{(A; \alpha)} = I_A \otimes \sigma_{\alpha} \otimes \sigma^{\mu}, \;  {\small A =0, 1,..,8; \; 
\alpha = 2,3; \; \mu = 0,1,2,3}\; .
\label{GammawithI}
\end{equation}
Let the $3 \times 3$ ``colour matrices" $I_A$ appearing as the first factor in (\ref{GammawithI}) be defined as follows:
$I_0 = {\mbox{l\hspace{-0.55em}1}}_{3}, \; I_r = Q_r, \; I_{r+3} = Q^{\dagger}_r, \; I_7 = B, \; I_8 = B^{\dagger}.$
Then the original $\Gamma$-matrices given by (\ref{Gammasbig}) are encoded as 
$\Gamma^0_{(8, 3)} = B^{\dagger} \otimes \sigma_3 \otimes {\mbox{l\hspace{-0.55em}1}}_{2}$
and $\Gamma^i_{(2; 2)} = Q_2 \otimes (i \sigma_2) \otimes \sigma^i$. The eight matrices with $A=1,2,...8$
with the multiplication rules given in (\ref{RelBQ}) span the ternary basis, generated by the cyclic $Z_3$-automorphism 
of the $SU(3)$ algebra (\cite{Kac1994}, Sect. $8$).

In order to get the closed formula for the action ${\cal{S}}^{(0)} \Gamma^{\mu} [{\cal{S}}^{(0)}]^{-1}$ of classical 
spinorial Lorentz symmetries generated by $L^{(0)}$, we should introduce the pairs of $\Gamma^{\mu}$-matrices 
$\Gamma^{\mu} = ( \Gamma^{i}_{(A ; 2)}, \; \Gamma^{0}_{(B;3)}) $ and $ {\tilde{\Gamma}}^{\mu} = 
(\Gamma^{i}_{(B; 2)}, \; \Gamma^{0}_{(A; 3)})$, $A \neq B$.
For any choice of $\Gamma^{\mu}$'s in (\ref{GammawithI}) we get:
\begin{equation}
\left[ J_i^{(0)}, \Gamma^{j}_{(A; \alpha)} \right] = \epsilon_{ijk} \Gamma^{k}_{(A; \alpha)}, \; \; \;  
\left[ J_i^{(0)}, \Gamma^{0}_{(A; \alpha)} \right] =  0,
\label{JGamma0}
\end{equation}
and the boosts $K^{(0)}_i$ act covariantly on doublets $\left( \Gamma^{\mu}, {\tilde{\Gamma}}^{\mu} \right)$ as follows: 
$$\left[ K_i^{(0)}, \Gamma^{j}_{(A; 2)} \right] = \delta^j_i \;  \Gamma^{0}_{(A; 3)}, \;  \; \;  
\left[ K_i^{(0)}, \Gamma^{0}_{(B; 3)} \right] = \Gamma^{i}_{(B; 2)},$$
%\left[ K_i^{(0)}, \Gamma^{j}_{(A; \alpha + 1)} \right] = \epsilon_{ijk} \Gamma^{k}_{(A; \alpha)}, $$
\begin{equation}
\left[ K_i^{(0)}, \Gamma^{j}_{(B; 2)} \right] = \delta_i^j \; \Gamma^{0}_{(B; 3)}, \; 
%\left[ K_i^{(0)}, \Gamma^{0}_{(A; \alpha)} \right] = \Gamma^{i}_{(A; \alpha + 1)}, \; \; \;  
\left[ K_i^{(0)}, \Gamma^{0}_{(A; 3)} \right] = \Gamma^{i}_{(A; 2)},
\label{Kongamma}
\end{equation}
(with $A \neq B)$, i.e. the standard Lorentz covariance requires the {\it doublet} of coloured Dirac spinors;  
%the second one using  the following partners of 
In particular, the $\Gamma^{\mu}$ matrices (\ref{Gammasbig}) should be supplemented by:
\begin{equation}
{\tilde{\Gamma}}^0 = \Gamma^{0}_{(2; 3)} = Q_2 \otimes (\sigma_3) \otimes {\mbox{l\hspace{-0.55em}1}}_{2},
\; \; \; {\tilde{\Gamma}}^i = \Gamma^{k}_{(8; 2)} = B^{\dagger} \otimes i \sigma_2 \otimes \sigma^{k}.
\label{Pairgammas}
\end{equation}
One can conjecture that the pairs of $\Gamma$-matrices generated by the standard Lorentz covariance requirement 
can be used fir the introduction of weak isospin doublets
of the $SU(2) \times U(1)$ electroweak symmetry. In such a way one can conclude that the internal symmetries
$SU(3) \times SU(2) \times U(1)$ of Standard Model follow from the imposition of standard Lorentz covariance
on colour Dirac multiplets.

Our next goal is to study the generalized Lorentz covariance of coloured Dirac equations, by generalization of 
standard invariance condition (\ref{Lorentz3}) and incorporating the standard $\Gamma^{\mu}$-matrices (\ref{Gammasbig}) 
into an irreducible representation of ${\cal{L}}$.
For this purpose, we should study the $18$-parameter symmetry transformation
$\Gamma^{\mu} \rightarrow {\cal{S}} \Gamma^{\mu} {\cal{S}}^{-1}$, where
\begin{equation}
{\cal{S}} = \prod_{r=0}^2 \; e^{ i \; [ \alpha^k_{(r)} J^{(r)}_k + \beta_{(r)}^m K^{(r)}_m ]},
\label{BigS}
\end{equation}
with $\alpha^k_{(0)}, \; \beta^k_{(0)}$ real, $(\alpha^k_{(1)})^{*} = \alpha^k_{(2)}, \; (\beta^k_{(1)})^{*} = \beta^k_{(2)},  \;
J_k^{\dagger (1)}  = J_k^{(2)}$ and $K_m^{\dagger (1)} = K_m^{(2)}.$
It follows that in order to obtain the closure of the faithful action of generators $(J_k^{(s)}, \; K_m^{(s)})$ ($s = 0,1,2$)
%so generalized spinorial transformations 
on matrices $\Gamma^{\mu}$,
one should introduce two sets $\Gamma^{\mu}_{(a)},  \Gamma^{\mu}_{\dot{a}} = (\Gamma^{\mu}_{(a)})^{\dagger}  \; (a = 1,2,...6)$ 
of coloured $12 \times 12$ Dirac matrices supplemented by Lorentz doublet partners $( {\tilde{\Gamma}}^{\mu}_{(a)}, {\tilde{\Gamma}}^{\mu}_{(\dot{a})}).$ 
If we choose $( J_k^{(1)}, K_m^{(1)} ) $ as given by Eqs. (\ref{J1K2}), (\ref{J2K1}), and assume that  $\Gamma^{\mu}_{(1)}$
is described by the formula (\ref{Gammasbig}), by calculating the multicommutators of $\left( J^{(1)}_i, K^{(1)}_l \right)$ with 
the set $\Gamma^{\mu}_{(a)}, \; (a=1,2...6)$,  we get the following sextet of $\Gamma$-matrices closed under the action of $L^{(1)}$ :
\begin{equation}
\begin{split}
&\Gamma^{\mu}_{(1)} = \left( \Gamma^0_{(8;3)}, \; \Gamma^i_{(2;2)} \right); \; \; 
\Gamma^{\mu}_{(2)} = \left( \Gamma^0_{(2;2)}, \; \Gamma^i_{(4;3)} \right); 
\\
&\Gamma^{\mu}_{(3)} =  \left( \Gamma^0_{(4;3)}, \; \Gamma^i_{(8;2)} \right); \; \;  
\Gamma^{\mu}_{(4)} = \left( \Gamma^0_{(8;2)}, \; \Gamma^i_{(2;3)} \right);
\\
%\begin{equation}
&\Gamma^{\mu}_{(5)} =  \left( \Gamma^0_{(2;3)}, \; \Gamma^i_{(4;2)} \right); \; \; 
\Gamma^{\mu}_{(6)} = \left( \Gamma^0_{(4;2)}, \; \Gamma^i_{(8;3)} \right).
\end{split}
\label{sixGammas}
\end{equation}
The realization of $L^{(2)}$ sector is obtained by introducing the Hermitean-conjugate sextet
 $\Gamma^\mu_{(\dot{a})} = (\Gamma^\mu_{(a)})^{\dagger}$; further one should add 
${\tilde{\Gamma}}^\mu_{(\dot{a})} = ({\tilde{\Gamma}}^\mu_{(a)})^{\dagger}$ due to standard Lorentz covariance.
The generalized Lorentz transformations of $24$ matrices  
 $\Gamma^{\mu}_{(F)} = ( \Gamma^{\mu}_{(a)}, \Gamma^{\mu}_{(\dot{a})}; \; {\tilde{\Gamma}}^{\mu}_{(a)}, {\tilde{\Gamma}}^{\mu}_{(\dot{a})} ) $ 
will be expressed by the following generalization of the formula (\ref{Lorentz3}) 
\begin{equation}
{\cal{S}} \Gamma^{\mu}_{(F)} {\cal{S}}^{-1} = \Lambda^{\mu \; (G)}_{\; \; \nu \; (F)} \; \Gamma^{\nu}_{(G)}, 
\; \; \; \mu, \nu = 0,1,2,3; \; \; \; F,G=1,2,...,24.
\label{SLambda}
\end{equation}
where with the help of the Baker-Campbell-Hausdorff type formula (\cite{Hall}) the matrix $\Lambda^{\mu \; (G)}_{\; \; \nu \; (F)} $
can be calculated explicitly if the multicommutators of $\Gamma^{\mu}_{(F)}$ with the generators of ${\cal{L}}$ are known
up to the sixth order (\cite{Greensite}).

In order to describe in compact way the action of generalized Lorentz algebra on coloured Dirac matrices,
we can introduce the $144 \times 144$ ``master" ${\bf{\Gamma}}^{\mu}$ matrices built up in a suitable manner as a $12 \times 12$
matrix with its entries being the $12 \times 12$ coloured $\Gamma^{\mu}_{(a)}$ 
defined in (\ref{sixGammas}), their Hermitean conjugates $\Gamma^\mu_{(\dot{a})}$ and their Lorentz doublet partners.. 
In such a way one can obtain the ``master" colour Dirac equation for $144$-component master spinor field describing
six known relativistic quarks in three flavour doublets $(u, d), \; (c, s), \; (t, b)$. In such a scheme
the sextet (\ref{sixGammas}) defining six colour multiplets introduces an additional $ Z_2 \times Z_3$ grading with the discrete degrees of freedom,
related with flavour doublets ($Z_2$ grading) and the three quark families, called also ``generations" ($Z_3$ grading).
At the present stage we assume that this second $Z_3$ grading, related with quarks' families, contrary to the colour $Z_3$ grading,
does not imply any entanglement of symmetries.
\vskip 0.3cm
\noindent
{\bf 4. Solutions, ternary products and confinement}} %\hskip 0.2cm 
\vskip 0.2cm
\indent
 Let us consider the solutions of the coloured Dirac equation (\ref{Dirac3})
%As in the case of any linear system, 
in the exponential form 
%whose Fourier images are  
$e^{k_{\mu} x^{\mu}}$.
The characteristic equation of the $12 \times 12$ operator 
(\ref{detGamma}) yields the dispersion relation
%$k_0^6 - \mid {\bf k } \mid^6 = m^6.$ 
in the $4$-momentum space of Fourier transforms:
\begin{equation}
k_0^6 - \mid {\bf k } \mid^6 = m^6, 
\label{dispersionk}
\end{equation} 
The general solution is therefore $e^{k_0 x^0 - {\bf k} \cdot {\bf r}}$, provided the above relation 
(\ref{dispersionk}) 
is satisfied, what for the choice of real $k_0$ means that it is given by
\begin{equation}
c k_0 = c \root 6\of{\mid {\bf k } \mid^6 + m^6 } =  \Omega ({\bf k}) .
%k_0 = \left[ \mid {\bf k } \mid^6 + m^6 \right]^{\frac{1}{6}} = c^{-1} \; \Omega ({\bf k} ),
\label{solvek0}
\end{equation}
Any sixth-order root of real number provides six different values, two real ones, ( $\pm \Omega, {\bf k}), \; {\bf k} $ real ), 
and four other ones obtained by multiplying $\pm \Omega ({\bf k} )$ by $j$ and $j^2$,
%the last two obtained by multiplication by $\pm j^2$, 
yielding  full set of six solutions
%%%*
with $ck_0$ given by  $\pm \Omega, \pm j \Omega, \pm j^2 \Omega.$
Further, if we have one solution with given $k_0$ and ${\bf k}$
satisfying (\ref{solvek0}), we get other solutions of the same form, with
${\bf k}$ replaced by $j {\bf k}$ or $j^2 {\bf k}$. 
(the change ${\bf k} \rightarrow - {\bf k}$ does not introduce independent solutions, 
because a three-vector ${\bf k}$  covers the entire sphere $S^2$; 
 $k_0 \rightarrow - k_0$ does matter as it distinguishes positive and negative energy states). 

Combining all these possibilities we arrive at $18$ different exponentials, $9$ with positive $k_0$ and $9$
with negative $k_0$. They can be organized in the following two sets of solutions 

\begin{equation}
\Psi^{+}_{(r,s)} (t, {\bf r}) = e^{j^r \Omega t + j^s {\bf k} \cdot {\bf r}}, \; \; \; 
\Psi^{-}_{(r,s)} (t, {\bf r}) = e^{-j^r \Omega t + j^s {\bf k} \cdot {\bf r}}, 
\label{settwopsi}
\end{equation}
where $s,r =0,1,2$ and $\Omega$ is given by (\ref{solvek0}).

The colour Dirac equation (\ref{Dirac3}) as a system of $12$ differential equations
 of first order should display only $12$ independent solutions, six with positive and six with negative frequencies.
We can choose the six off-diagonal entries in (\ref{settwopsi}), with $r \neq s$, which can be displayed in the
following matrix:
\begin{equation}
\begin{pmatrix}  0 & e^{\Omega\,t + j {\bf k \cdot r}}
& e^{\Omega\,t + j^2 {\bf k \cdot r}} \cr 
e^{j \Omega\,t + {\bf k \cdot r }} & 0 & e^{j \Omega\,t + j^2 {\bf k \cdot r}} \cr
e^{j^2 \Omega\,t + {\bf k \cdot r}}& e^{j^2 \Omega\,t + {\bf k \cdot r}}  & 0  
\end{pmatrix}, 
\label{sixmatrix}
\end{equation}
and similarly for the negative energy values $(\Omega \rightarrow - \Omega)$; one can check that
the determinant of the matrix (\ref{sixmatrix}) displaying the $6$ independent solutions is equal to $1$.

All these twelve functions, describing propagation of coloured quarks, 
do not represent free waves due to the presence of damping factors. However, 
observing that there are only two ways of obtaining imaginary units as linear combinations of the $Z_3$ roots $1, \; j, \; j^2,$
namely
\begin{equation}
\frac{1}{\sqrt{3}} (1 +2 j) = i , \; \; \; \frac{1}{\sqrt{3}} (1 + 2 j^2) = - i 
\label{rootsi}
\end{equation}
one can produce propagating free wave-like solutions by forming two independent cubic products with positive $\Omega$, 
and two ones with negative $\Omega$.
Following (\ref{sixmatrix}), (\ref{rootsi}) we choose the first pair of solutions as
\begin{equation}
\Psi^{+}_{(B)} (t, {\bf r}) = \Psi^{+}_{(2,0)} \Psi^{+}_{(0,1)} \Psi^{+}_{(2,1)}  = e^{- i \sqrt{3} (\Omega t + {\bf k}{\bf r})}, 
\label{TwoFGH1}
\end{equation}
\begin{equation}
 {\bar{\Psi}}^{+}_{(B)} (t, {\bf r}) = \Psi^{+}_{(1,0)} \Psi^{+}_{(0,2)} \Psi^{+}_{(1,2)} = e^{i \sqrt{3} (\Omega t + {\bf k}{\bf r})},
\label{TwoFGH2}
\end{equation}
and two ones with negative $\Omega$.
With two additional solutions obtained by replacing $\Omega$ by $-\Omega$ we get just the right number of four plane wave solutions needed to describe
a massive spin $\frac{1}{2}$ particle - a composite three-quark free baryon wave function. Similarly, due to the relation $(j - j^2 )/{\sqrt{3}}=i$,
the quark-antiquark pairs of solutions with positive and negative frequencies will provide the particle and anti-particle spin-0 meson plane waves.

The $4$-vector $[\frac{\Omega}{c}, {\bf k}]$ in baryonic wave functions (\ref{TwoFGH1}, \ref{TwoFGH2}) does not satisfy the usual 
quadratic dispersion relation $\omega^2 = c^2 {\bf k}^2 + m^2$, where $m$ is the baryonic mass, but the relation (\ref{dispersionk}),
i.e. $\Omega^6 = c^6 \mid {\bf k} \mid^6 + M^6$. One can argue however that because for $\mid {\bf{k}} \mid \gg M$ we have
\begin{equation}
\Omega^2 = \root 3\of{ c^6 ( {\bf{k}}^2 )^3 + M^6} = c^2 {\bf{k}}^2 \root 3\of{1 + \frac{M^6}{\mid {\bf{k}} \mid^6} } \simeq
c^2 {\bf{k}}^2 + m^2 ( {\bf{k}}),
\label{Omegasq}
\end{equation}  
where 
\begin{equation}
m^2 ( {\ bf{k}}) = \frac{1}{3} M^2 \left[ \frac{M^4}{c^4 {\bf{k}}^4} + {\cal{O}} \left( \frac{M^2}{c^2 {\bf{k}}^2} \right)^5 \right],
\label{msquare}
\end{equation}
the baryonic wave functons (\ref{TwoFGH1}, \ref{TwoFGH2}) satisfy the d'Alembert equation with source term which
quickly converges on the solutions  (\ref{TwoFGH1}, \ref{TwoFGH2}) to zero in the high energy limit ${\bf{k}} \rightarrow \infty$.

An important future task is to construct a QCD framework with colour Dirac spinors.
The  presented ideas are preliminary; in principle it should be possible to introduce the generalized Dirac action incorporating 
the ``master'' colour Dirac matrices which could describe all phenomenologically known quarks.
\vskip 0.3cm
\indent
{\bf Acknowledgements} \; Many interesting and useful discussions with Jan-Willem van Holten
and Michel Dubois-Violette, 
as well as valuable comments by Piotr Kosi\'{n}ski and Andrzej Sitarz are gratefully acknowledged.
 J.L. has been supported by Polish National Centre (NCN) project  2017/27/B/ST2/01902 and by COST action
MP 1405 QSPACE.

%%%#########################

\end{document}